\documentstyle[aps,epsf,twocolumn]{revtex}

\begin{document}

\title{\bf 
Conditional Dynamics of Open Quantum Systems: \\
the Case of Multiple Observers
}
\author{ Jacek Dziarmaga$^{1,2}$, 
         Diego A. R. Dalvit$^1$ ,
         and 
         Wojciech H. Zurek$^1$ }
\address{ 1) Los Alamos National Laboratory, T-6, Theoretical Division, 
             MS-B288, Los Alamos, New Mexico 87545, USA \\
          2) Instytut Fizyki Uniwersytetu Jagiello\'nskiego, 
             Reymonta 4, 30-059 Krak\'ow, Poland
\date{April 16, 2001}
}
\maketitle

\begin{abstract} 
{\bf Many observers can simultaneously measure different parts of 
an environment of a quantum system in order to find out its state.
To study this problem we generalize the formalism of conditional 
master equations to the multiple observer case. To settle
some issues of principle which arise in this context 
(as the state of the system and of the environment are
ultimately correlated), we consider an example of a system 
qubit interacting through controlled nots (CNOTs) with  environmental
qubits. The state of the system is the easiest to find out for observers
who measure in a basis of the environment which is most 
correlated with the pointer basis of the system. In this case 
the observers agree the most. Furthermore, the more predictable 
the pointers are, the easier it is to find the state of the system, 
and the better is the agreement between different observers. }
\end{abstract}


%%%%%%%%%%%%%%%%%%%%%%%%%%%%%%%%%%%%%%%%%%%%%%%%%%%%%%%%%%%%%%%%%%%

Pointer states are the states of a system which get entangled 
the least with the environment. They are therefore the most predictable 
and, hence, the most classical states of the system
\cite{pointer,zurek,zhp}. In our recent Letter \cite{us} we 
considered an observer performing continuous quantum 
measurements on an environment of a system in order to monitor 
and predict its state \cite{carmichael,wiseman1}. 
We found that under reasonable assumptions pointer states of
the system do not depend on the basis selected by the
observer to carry out measurements on the environment.
We also found evidence that measurements in a basis of the 
environment which is most strongly correlated with the system 
are most efficient in yielding information about its state.

In this Letter we consider several observers monitoring different 
parts of the environment to extract information about the system.
We shall show that, again, each observer gains most information
from measurements in a basis which is most strongly correlated 
with the pointer states. However, in the presence
of multiple observers new questions arise about correlations 
between the state different observers ascribe to the system.
We find that when all observers measure their 
environments in a basis correlated to the pointer states,  
then the indications of their appartuses are very strongly 
correlated, as might have been expected for measurements of 
very classical states. On the other hand, when observers 
measure in a basis poorly correlated to the pointer states
or when the preferred pointer states are not very classical,
then it is possible that apparatuses disagree for large
fraction of time (see also \cite{anals} for an information
theoretic discussion of related issues).

Many-body entanglement occurs
in course of decoherence when several
subsystems of the environment get entangled with
the system. For instance, a one-qubit system and, say, 
two one-qubit environments
can find themselves in a GHZ-like state 

\begin{equation}\label{GHZ}
\frac{1}{\sqrt{2}} \left(
|1\rangle_{\rm{S}}
|1\rangle_{\rm{E}_1}
|1\rangle_{\rm{E}_2}
\;+\;
|0\rangle_{\rm{S}}
|0\rangle_{\rm{E}_1}
|0\rangle_{\rm{E}_2} \right) \;\;.
\end{equation}
We shall focus on this ideal case, as it allows us to illustrate
interesting issues of fundamental importance that arise in the case
of multiple observers. 

The reduced density matrix of the system is mixed, 
$\rho=(|1\rangle\langle 1|+|0\rangle\langle 0|)/2$.
Imagine that environments $\rm{E}_1$ and $\rm{E}_2$ are measured 
by different observers $1$ and $2$ who know beforehand that 
the total state is (\ref{GHZ}). Observer $1$ measures the 
state of $\rm{E}_1$ in the $\{|1\rangle,|0\rangle\}$ basis. 
If his measurement result is $|1 \rangle (|0\rangle)$, then 
he discovers that the system (and, by the way, the 
other environment) are in the state $|1\rangle(|0\rangle)$. 
If his measurement is followed by the measurement of  
observer $2$, then observer $2$ will also find both 
his environment and the system in the $|1\rangle (|0\rangle)$ 
state. The $\{ |1\rangle, |0\rangle\}$ basis is a good choice 
in the sense that each observer alone can find out about the
system state. 

Suppose that the observers want to find out about the state of the system
in another basis, say the Hadamard transformed 
basis $|\pm\rangle =(|1\rangle\pm|0\rangle)/\sqrt{2}$.
Suppose that observer $1$ made a measurement in this basis 
and that his outcome is ``$+$". His measurement projects 
GHZ state (\ref{GHZ}) onto
$|1\rangle_{\rm{S}} |+\rangle_{\rm{E}_1}
|1\rangle_{\rm{E}_2}
\;+\; |0\rangle_{\rm{S}} |+\rangle_{\rm{E}_1}
|0\rangle_{\rm{E}_2}$.
The reduced density matrix of the system is not affected at all: 
it remains in the initial mixed state
as before the measurement. The single observer can find nothing 
about the state of the system when he measures in the ``wrong" 
$\{|+\rangle,|-\rangle\}$ basis. Let observer 2 step in and make 
his measurement. If he 
measures in the ``good'' basis, then he gets full information 
about the system. If, on the contrary, he measures in the 
``wrong'' basis, alone he will not be able to ascertain the 
state of the system. Suppose that the result of his measurement 
is, say, ``$-$". Then the previous state is further projected on
$|-\rangle_{\rm{S}} |+\rangle_{\rm{E}_1}
|-\rangle_{\rm{E}_2} $, 
and the system is in the pure ``$-$" state,
$\rho=|-\rangle\langle-|$. Having measured $\rm{E}_1$ 
and $\rm{E}_2$ in the $\{|+\rangle,|-\rangle\}$ basis, each 
observer alone is ignorant of the state of the system.
However, correlations between their 
measurement results contain full information about the system. 
If the outcomes of the two observers are ``$++$" or
``$--$" then the system state is ``$+$", but if the outcomes are
``$+-$" or ``$-+$" then the system state is ``$-$".

%%%%%%%%%%%%%%%%%%%%%%%%%%%%%%%%%%%%%%%%%%%%%%%%%%%%%%%%%%%%%%%%%%%

In the GHZ example above the state (\ref{GHZ}) was known 
to both observers beforehand. Given that knowledge, and
after a fortuitous choice of the observables, they could 
draw unambiguous conclusions about the state of the system after just 
one projection. In practice correlations between an unknown
state of the system and different parts of the environment arise 
as a result of interaction. Let us consider a toy 
example that illustrates such a scenario. Let the system be 
a single qubit $\cal S$ with zero self-Hamiltonian.
It is initially prepared in a state

\begin{equation}
\rho^{t_0}=
\sum_{a,b=0,1}
\rho^{t_0}_{ab}\;
|a\rangle\langle b|\;\;.
\end{equation}
Let the environment of such a qubit be an ensemble of pairs of
qubits, all initially prepared in state $|0\rangle$ 
($|1\rangle$ and $|0 \rangle$ are eigenstates of $\sigma_z$ 
with eigenvalues $+1$ and $-1$ respectively). We want to find 
out the state of the system from measurements on the
environment. The environmental qubits entangle with the system as 
shown in Fig.1. At the time $t_1$ the first pair of environmental
qubits denoted by $(1,t_1)$ and $(2,t_1)$ is put in contact
with the system. The system acts as a control on the environmental
qubits, perfoming a C-NOT operation on both of them, so that 
the system qubit and the environmental pair of qubits get fully 
entangled. After completion of these operations the first pair is 
decoupled from the system. At the time $t_2$ a second pair is entangled with 
the system in a similar way. After $n$ such double CNOT 
operations the total density matrix becomes

\begin{eqnarray}
\label{rhonSE}
\rho^{t_n}_{\cal S+E} &=&
\sum_{a,b=0,1}
\rho^{t_0}_{ab}\;
|a\rangle\langle b|\;\otimes\;
|a\rangle\langle b|_{(1,t_1)}\;\otimes\;
|a\rangle\langle b|_{(2,t_1)} \nonumber \\
&& \otimes \dots\otimes\;
|a\rangle\langle b|_{(1,t_n)}\;\otimes
|a\rangle\langle b|_{(2,t_n)}\;.
\end{eqnarray}
The reduced density matrix of the system after $n$ steps,
$\rho^{t_n}$, can be obtained from $\rho^{t_n}_{\cal S+E}$ by 
tracing over the environment, $\rho^{t_n}=\;
{\rm Tr}_{\cal E }\;\rho^{t_n}_{\cal S + E }\;=\;
\rho_{00}^{t_0} \; | 0\rangle\langle 0 | \;+\;
\rho_{11}^{t_0} \; | 1 \rangle\langle 1 | $
for any $n>0$. It becomes diagonal in the $\{|0\rangle,|1\rangle\}$
basis already after the entanglement with the first pair.
In the steps that follow $\rho^{t_n}$ does not change any more. 
This description can be encapsulated in the following difference 
equation for the matrix elements

\begin{equation}\label{UMEtoy}
\rho_{ab}^{t_n} \;=\; \delta_{ab} \;\rho_{ab}^{t_{n-1}} \;.
\end{equation}
This is a markovian master equation, as the
next state of the system depends only on its immediate
predecesor (and not on the history). Since it was obtained 
by tracing out the environment, it is an ``unconditional'' 
master equation (UME), in the sense that all the information 
about the environment is ignored. If the system is initially 
in one of the states $|0\rangle$ or $|1\rangle$, then the 
state of the system does not get entangled with the environment and 
it does not lose any purity. In other words, the states 
$|0\rangle$ and $|1\rangle$ are perfect pointer states.

Suppose that the information about the state of $\cal E$ 
is not ignored. Let there be two observers labelled $\alpha=1,2$ 
who measure the operators 
$\hat{\sigma}_{\alpha}=x_{\alpha}\sigma_x+
y_{\alpha}\sigma_y+z_{\alpha}\sigma_z$, 
where $x_{\alpha}^2+y_{\alpha}^2+z_{\alpha}^2=1$,
on the environmental qubits $(\alpha,t_n)$.
We shall denote the state of the environmental qubit $\alpha$ after the 
measurement as $|N^{t_n}_{\alpha}\rangle$, where $N^{t_n}_{\alpha}$ (which
can be either $+1$ or $-1$) is the result of the measurement.
Since the $\sigma_z$ basis is the one correlated with the pointer states,
we expect that observers measuring in that basis will most efficiently
gain information and agree the most about the state of the system \cite{us}.

\begin{figure}[h]
\centering \leavevmode \epsfxsize=8cm
\epsfbox{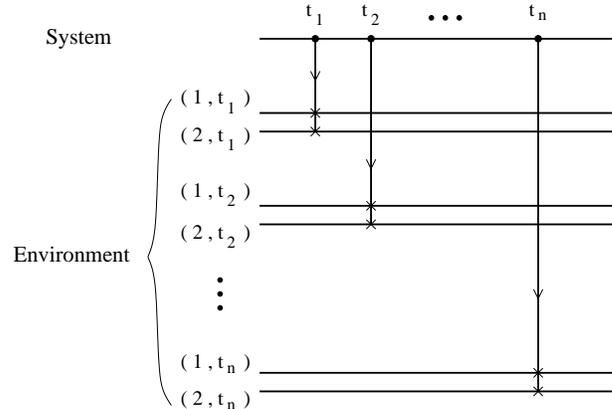} 
\caption{C-NOT circuit for the toy model. Each pair 
$(\alpha,t_n)$ of qubits of the environment (target qubits) 
interacts only once with the system (control qubit). We recall 
that the logical operation C-NOT flips the state of the target if the state 
of the control is 1, and does nothing otherwise. }
\end{figure}

To make the measurement on an environmental qubit the observer fully 
entangles it with his measuring apparatus (in effect, using the environmental
qubit as a control to perform CNOT in an eigenbasis of 
$\hat{\sigma}_{\alpha}$) and then the state of the memory 
decoheres in its pointer basis. The reduced density 
matrix of the memory qubit becomes diagonal in the measurement basis. 
Measurements on environmental qubits can be carried out at any time 
after they have interacted with the system. Their results affect the 
observers'  knowledge about the state of the system at time 
$t_n$ when the $n$-th pair of qubits got entangled with the system. This 
knowledge is expressed by the reduced density matrix of the 
system $\rho^{t_n}$. From the point of view of $\rho^{t_n}$ 
this whole arbitrarily delayed measurement process can be 
described by the projections of the full $\rho_{\cal S+E}^{t_n}$ 
in the measurement basis of the $n$-th pair of qubits. 

After $n-1$ steps followed by $2(n-1)$ measurements with
the outcomes $\{ N_{\alpha}^{t_n} \}$ in the measurement basis of
observers $\alpha=1,2$, the density matrix of the system and
the environment, conditioned on given set of measurement records, is

\begin{eqnarray}
&&\rho_{\cal S + E}^{t_n} =
|N_1^{t_1}\rangle\langle N_1^{t_1}| \;\otimes
\dots\otimes\;
|N_2^{t_{n-1}} \rangle\langle N_2^{t_{n-1}}|  \\
&& \otimes
\left(
\sum_{a,b=1,0}
\rho_{ab}^{t_{n-1}}\;
|a\rangle \langle b| \;\otimes\;
|a\rangle\langle b|_{(1,t_n)} \;\otimes\;
|a\rangle\langle b|_{(2,t_n)} 
\right) , \nonumber
\end{eqnarray}
where $\rho^{t_{n-1}}$ is the reduced density matrix of the
system after $n-1$ steps conditioned on the measurement
results at the times $t_1,\dots,t_{n-1}$. The unnormalized conditional
reduced density matrix of the system becomes

\begin{eqnarray}\label{tilderho}
\tilde{\rho}^{t_n} &=&
{\rm Tr}_{{\cal E}-{\cal E}_n}\; 
\langle N_1^{t_n} |\; \langle N_2^{t_n} |\;
\rho_{\cal S + E}^{t_n}\; | N_1^{t_n} \rangle\; | N_2^{t_n} \rangle\; 
\nonumber \\ 
&=& \frac{1}{4} \;\rho_{11}^{t_{n-1}}\; |1\rangle\langle1|\; 
(1+z_1 N_1^{t_n})(1+z_2 N_2^{t_n})  \nonumber \\
&+& 
\frac{1}{4} \;\rho_{00}^{t_{n-1}} \; |0\rangle\langle 0|\; 
(1-z_1 N_1^{t_n})(1-z_2 N_2^{t_n}) \nonumber\\
&+ & \frac{1}{4} \;N_1^{t_n} \; N_2^{t_n} \;\rho_{10}^{t_{n-1}} \; 
|1\rangle\langle0|\;  \sqrt{1-z_1^2}\sqrt{1-z_2^2} \nonumber \\
&+& 
\frac{1}{4} \;N_1^{t_n}\;N_2^{t_n} \;\rho_{01}^{t_{n-1}}\; 
|0\rangle\langle1|\; \sqrt{1-z_1^2} \sqrt{1-z_2^2}\;\;.
\end{eqnarray}
The probability to get a given outcome $(N_2^{t_n}, N_2^{t_n})$
and the normalized reduced density matrix of the system
conditioned on this outcome are

\begin{eqnarray}
P(N_1^{t_n},N_2^{t_n} | \rho^{t_{n-1}}) &=&
{\rm Tr} \;\tilde{\rho}^{t_n} \label{MOCMEprob} \\
&=&
\frac14\;\rho_{11}^{t_{n-1}} \;(1+z_1 N_1^{t_n}) (1+z_2 N_2^{t_n} ) 
\nonumber \\  
&+& \frac14\;\rho_{00}^{t_{n-1}} \;(1-z_1 N_1^{t_n}) (1-z_2 N_2^{t_n} ) 
; \nonumber \\
\rho^{t_n} &=&
\frac{\tilde{\rho}^{t_n}}{P(N_1^{t_n},N_2^{t_n} | \rho^{t_{n-1}})} \;.
\label{MOCMEtoy}
\end{eqnarray}
The above equation is a multiple observer
conditional master equation (MOCME). It describes
the evolution of the knowledge about the state of the system $\rho^{t_n}$ of 
a ``supervisor'' who has access to the measurement records 
$N_{\alpha}^{t_n}$ of all the observers.
The average of the conditional $\rho^{t_n}$ over
different outcomes $N_{\alpha}^{t_n}$ weighted by their
probability distribution $P(N_1^{t_n},N_2^{t_n} | \rho^{t_{n-1}})$
gives the unconditional master equation (\ref{UMEtoy}).

Suppose that both observers measure in the $\sigma_z$ basis.
The probability distribution 
$P(N_1^{t_n},N_2^{t_n} | \rho^{t_{n-1}})=
\delta_{N^{t_n}_1,N^{t_n}_2}\;
(\rho^{t_{n-1}}_{11}\;\delta_{N^{t_n}_1,+1}\;+\; 
\rho^{t_{n-1}}_{00}\;\delta_{N^{t_n}_1,-1})$
implies that the results of the two observers
are completely correlated, $N^{t_n}_1=N^{t_n}_2$.
The full correlation follows from the entanglement
between the system and the environmental qubits. The result
$N^{t_n}_1=N^{t_n}_2=+1 (-1)$ obtained with the probability
$\rho^{t_{n-1}}_{11} (\rho^{t_{n-1}}_{00})$ gives a conditional 
$\rho^{t_n}=|1\rangle\langle 1|  (\rho^{t_n}=|0\rangle\langle 0|) $. 
In the $\sigma_z$-basis  it is enough to measure just one 
environmental qubit to gain full knowledge about the state of the system 
(and purify $\rho^{t_n}$). The fully correlated observers
always agree that this state is $|1\rangle$ or $|0\rangle$. We note again 
that a weighted average over the results $N^{t_n}_{\alpha}$ gives 
the unconditional
$\rho^{t_n}=\;\rho^{t_{n-1}}_{11}\;|1\rangle\langle1|\;+\;
\rho^{t_{n-1}}_{00}\;|0\rangle\langle0|$ in agreement
with the UME (\ref{UMEtoy}). 
  
It is not so easy to find the state of the system when the observers 
measure in a basis which is less well correlated to the 
pointer states. To illustrate this we take for definiteness 
$z_1=z_2=\epsilon$ with $0\leq\epsilon\ll 1$. We will
follow the evolution of the observers' knowledge about the state of the system 
with the help of the polarization 
$A^{t_n}=\rho^{t_n}_{11}-\rho^{t_n}_{00}$ they infer from their
measurements. Full knowledge of a predictable pointer state, a pure 
$|0\rangle$ or $|1\rangle$ state of the system, 
corresponds to $A^{t_n}=\pm 1$. We use
the MOCME (\ref{MOCMEtoy}) and expand to leading order
in $\epsilon$ to derive a stochastic master equation for $A^{t_n}$,

\begin{eqnarray}
\frac{ A^{t_n}-A^{t_{n-1}} }{ \epsilon } &=&
(N^{t_n}_1+N^{t_n}_2)[1-(A^{t_{n-1}})^2]\;, \nonumber\\
P(N^{t_n}_1,N^{t_n}_2 | A^{t_{n-1}}) &=&
\frac14\;[\;1\;+\;\epsilon(N_1^{t_n}+N_2^{t_n})A^{t_{n-1}}\;]\;.
\end{eqnarray}
Measurements of $\sigma_z$ yield $A^{t_n}=\pm 1$, that are fixed points of 
this equation. For $\epsilon=0$
(e.g. $\sigma_x$-measurement) the polarization $A^{t_n}$ 
does not change at all: if we start from $-1<A^{t_0}<+1$, then using such
$\epsilon=0$ measurements observers will never find 
out whether $A^{t_n}$ is $+1$ or $-1$. The outcomes of their measurements
do not depend on the state of the system. 
For $0<\epsilon\ll 1$ patient observers will find out the state of
the system if they measure $\propto 1/\epsilon^2$ environment.
The polarization $A^{t_n}$ makes a random walk. When $A^{t_n}$ walks into the 
area $A>0$, then in the next measurement the sum
$N^{t_n}_1+N^{t_n}_2$ will more likely come out positive than
negative and it will probably drive $A^{t_n}$ to be even more
positive. Eventually, after $\approx 1/\epsilon^2$ environmental
qubits get entangled with the system, $A^{t_n}$ will settle at $+1$ or at $-1$.
The closer is the measurement basis correlated with
the pointer states, the faster it is to find the state of the system.

So far we have described the evolution of $\rho^{t_n}$ as if we 
knew the records of both observers. This is rarely the case. Suppose that
observer $1$ knows only his own records $N^{t_n}_1$.
What density matrix $\rho^{t_n}_1$ represents his knowledge about the 
state of the system? Since he does not know $N^{t_n}_2$ the 
best he can do is to treat the other observer as if he were an environment, 
i.e.,  assume for $N^{t_n}_2$ a probability 
distribution like in Eq.(\ref{MOCMEprob}) and average 
the ${\tilde{\rho}}^{t_n}$ in Eq.(\ref{tilderho}) over $N^{t_n}_2$ with
this distribution. The weighted average is
$\tilde{\rho}^{t_n}_1 =
(\rho^{t_{n-1}}_{11}\;|1\rangle\langle1|\;(1+z_1 N^{t_n}_1)
+ \rho^{t_{n-1}}_{00}\;|0\rangle\langle0|\;(1-z_1 N^{t_n}_1))/2$.
The right hand side (RHS) of this equation still depends 
on the multiple observer $\rho^{t_{n-1}}$ but the observer $1$ 
does not know $\rho^{t_{n-1}}$ because he does not know any
earlier records of observer 2. In this situation the
best he can do is to take an average of the RHS over
the earlier records of observer 2:
$N^{t_{n-1}}_2,N^{t_{n-2}}_2,\dots,N^{t_1}_2$. By definition,
this average replaces $\rho^{t_n}$ on the RHS by the single 
observer density matrix $\rho^{t_n}_1$. After normalisation, so
that ${\rm{Tr}}\rho^{t_n}_1=1$, we obtain a single
observer conditional master equation (SOCME) for the observer $\alpha=1$,
conditioned only on his own records $N^{t_n}_{\alpha}$,

\begin{equation}\label{SOCMEtoy}
\rho_{\alpha}^{t_n}=
\frac{ 
\rho^{t_{n-1}}_{\alpha,11}\; 
|1\rangle\langle1|\; 
(1+z_{\alpha} N^{t_n}_{\alpha})\;+\;  
\rho^{t_{n-1}}_{\alpha,00}\; 
|0\rangle\langle0|\; 
(1-z_{\alpha} N^{t_n}_{\alpha})\; 
} 
{ 
\rho^{t_{n-1}}_{\alpha,11}\;  
(1+z_{\alpha} N^{t_n}_{\alpha})\;+\;  
\rho^{t_{n-1}}_{\alpha,00}\;  
(1-z_{\alpha} N^{t_n}_{\alpha}) 
} \;\;.
\end{equation}
The SOCME gives us a tool to check if and how fast do the observers $1$
and $2$ reach agreement about the state of the system. To this end we
define ``single observer" polarizations
$A^{t_n}_{\alpha}=\rho^{t_n}_{\alpha,11}-\rho^{t_n}_{\alpha,00}$. We use
Eq.(\ref{SOCMEtoy}) and an expansion to leading order in $\epsilon$ to
derive a stochastic equation for $A^{t_n}_{\alpha}$ conditioned on
$N^{t_n}_{\alpha}$,

\begin{equation}
\frac{ A^{t_n}_{\alpha}-A^{t_{n-1}}_{\alpha} }{ \epsilon }=\;
N^{t_n}_{\alpha}\;[1-(A^{t_{n-1}}_{\alpha})^2]\;.
\end{equation}
If the supervisor's polarization $A^{t_n}$ finally settles at $\pm1$, then
the probability distribution in Eq.(\ref{MOCMEprob})  will prefer positive
values for both $N^{t_n}_{\alpha}$'s and both $A^{t_n}_{\alpha}$'s will
follow $A^{t_n}$ to $\pm1$ (see Fig. 2). If the observers finally meet and
compare their results, they will fully agree.  In order to get this
agreement it is needed to entangle with the system an amount $\propto
1/\epsilon^2$ of pairs of environmental qubits, which is the smaller the
better are the measured states of the environment correlated to the
pointer states.

We derived the SOCME by averaging over the unknown records of the other
observer.  Note that the probability distribution in (\ref{MOCMEtoy})  is
linear in $\rho^{t_{n-1}}$ and can be easily averaged over the earlier
records of the observer $2$ at the earlier times
$t_{n-1},t_{n-2},\dots,t_0$. As a result the $\rho^{t_n}$ in the
distribution (\ref{MOCMEtoy}) averages to $\rho^{t_{n-1}}_1$. Such a
partially averaged distribution can be traced over $N^{t_n}_2$ to give a
distribution for a single observer $\alpha=1$,

\begin{eqnarray}
P_{\alpha}(N^{t_n}_{\alpha} | \rho^{t_{n-1}}_{\alpha}) &=& 
\frac12\;\rho^{t_{n-1}}_{\alpha,11}\;(1+z_{\alpha} N^{t_n}_{\alpha}) 
\nonumber \\
&+&   
\frac12\;\rho^{t_{n-1}}_{\alpha,00}\;(1-z_{\alpha} N^{t_n}_{\alpha})\;\;.
\label{Palphatoy}
\end{eqnarray}   
Eqs. (\ref{SOCMEtoy},\ref{Palphatoy})  can be regarded as a single observer
stochastic generator for the string of records $\{ N^{t_n}_{\alpha}\}$.
The multiple observer conditonal master equation (\ref{MOCMEtoy})  can
also be regarded as a ``multiple observer" stochastic generator for the
two strings of records $\{ N^{t_n}_1,N^{t_n}_2 \}$. From our derivation it
is clear that if we are given just one string, say, $\{ N^{t_n}_1 \}$,
then we will not be able to find out if the string comes from the multiple
observer or from the single observer generator. If records were not
independent, causality would be in trouble. 
One could send the entangled qubit $2$ to a distant
galaxy where the observer $2$ would make his measurements. By choosing to
measure or not to measure, or by changing the measurement basis, 2 could
affect records $N^{t_n}_1$ of the other observer and could signal with
superluminal velocity or backwards in time.  While causality is a general
and fundamental requirement, record independence is assured by the nature
of our environment. Each qubit pair is put in contact with the system only
once. Once they decouple they can no longer perturb the system state. On
the other hand, if they did not decouple, then in general it would be
possible to find out what observer 2 is doing without violating causality.

\begin{figure}
\centering \leavevmode \epsfxsize=9cm
\epsfbox{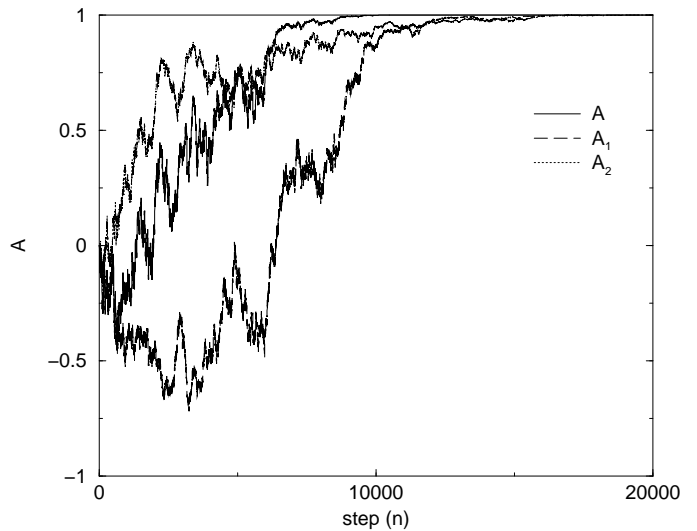} 
\caption{A single realization of the stochastic trajectories for the
polarizations $A^{t_n}$, $A^{t_n}_1$ and $A^{t_n}_2$. The initial condition
for all of them is null polarization. In this figure $\epsilon=10^{-2}$.} 
\end{figure}

In conclusion, we have shown that when several observers perform measurements
on the environment of a system, they agree most about
the state of the system if their measurement basis are correlated with the 
pointer states. For any other measurement basis their gain of information
is less efficient, and they can even gain no information at all if they
choose a ``wrong'' measurement basis. These results can be generalized to
the more realistic (but also more cumbersome) case of continuous quantum
measurement \cite{future}.

We are grateful to Juan Pablo Paz for discussions. 
This research was supported in part by NSA.

%%%%%%%%%%%%%%%%%%%%%%%%%%%%%%%%%%%%%%%%%%%%%%%%%%%%%%%%%%%%%%%%%%%

\end{document}